\documentclass{ws-procs975x65}
\usepackage{epsfig}
\begin{document}

\title{Numerical simulation of general relativistic stellar collapse}

\author{Cristi\'an R. Ghezzi and Patricio S. Letelier}

\address{Instituto de Matem\'atica Estat\'{\i}stica e Computa\c c\~ao Cient\'{\i}fica \\
Universidade Estadual de Campinas (UNICAMP-IMECC), \\
Cidade Univesit\'aria, Bar\~ao Geraldo, 13083-859,\\
Campinas, S\~ao Paulo, Brazil\\ 
E-mail: ghezzi@ime.unicamp.br}


\maketitle

\abstracts{
We present preliminar results and tests of a new general relativistic code to simulate
the hydrodynamic collapse of a $21$ solar masses star. We have assumed spherical symmetry
and used the formalism of Misner and Sharp to construct a finite-difference scheme
to solve the  Einstein's equations, energy-momentum conservation equations
and baryonic conservation equation. The code is similar to the one originally developed by
May and White (1967).
Here we discuss the capabilities of the code that make it well suited for numerical relativity
on a personal computer and some caveats
based on the experiments we have made with it.
}

\section{Introduction}

In this work we have reproduced the code of May and White (1967) (hereafter MW) using the same 
numerical techinques (with some little modifications) known
from their pioneering work. We had also tested  some modifications
suggested by Van Riper (1979) (hereafter VR). 
The output of our code can be used as an initial Cauchy data for more sophisticated relativistic codes 
(see for example Baumgarte et al. 1995).   
The code developed here altough based on old numerical techniques
is simple to implement and little time consuming, which
make it best suited for general relativistic simulations on a single personal computer.
Here we briefly resume the numerical caveats on a simulation of the hydrodynamic collapse of
a 21 solar masses star that forms a black hole. The complete results of astrophysical interest
will be published elsewhere (Ghezzi \& Letelier 2003).

\section{Numerical methods and initial conditions}
 We initially assumed uniform
density perfect fluid balls, with adiabatic equation of state $P=K \rho^{\gamma}$,
where $\gamma$ is the adiabatic coefficient which is set equal to $5/3$. It 
 is also possivel to implement other equations of state in our code.
The initial configuration resembles a newtonian star or a newtonian star-core collapsing under
the gravity when nuclear fuel is exhausted. During the time evolution a strong field regime
is achieved during which a neutron star or a black hole (a trapped surface) may be formed 
depending on the initial conditions.
The formation of the black hole was checked by the collapse of the $g_{tt}$ metric 
coefficient (collapse of the ``lapse function''). 
Alternatively, the black hole formation could be checked  by observing the convergence 
of the ligth ray paths that pass near a Lagrangian 
coordinate $\mu$, such that $2 G M(\mu)/R c^2 \ge 1$.

 It is possivel to assume arbitrary density or energy
profiles as initial data sets, altough we must observe that could be troublesome 
if discontinuities are present on the initial data.

We have used a  Lax-Wendroff scheme (two step Ritchmeyer's version) 
coupled with a  Crank-Nicolson (1947) algorithm to solve the
 system of equations (similar to the techniques described in MW, and in VR).
We have made tests using the viscocities given in VR and in MW.
The boundary conditions are that the pressure is zero at surface, the 
mass of the star is zero at the centre of the star,
and the coefficient $g_{tt}$ of
the metric is equal to one at the surface of the star. This last boundary condition
assures that the coordinate time are synchronised with the clock of a comoving observer
at the surface of the collapsing star.
The results obtained from one run test are shown in the figures (a)-(d).

A novelty feature of our code is that it don't break down when a black
hole forms. This is because
the code was built to self-adjust the time step in order to get the desired accuracy, and
eventually the time step becames exceedingly small when a black hole is obtained as the final
outcome of the simulation.
The  code is very fast and accurate.

\section{Discussion and Conclusion}

Using the VR viscocity or the MW viscocity seems to give no different
results using a low number of radial zones ($\sim 70$). However, when a larger 
number of zones is set the MW viscocity seems to work better if it is 
multiplied by some constant factor greater than one. This is necessary to enlarger the shock
waves through the zones, because when a large number
of radial zones is used the shocks have a steeper representation, and this in turns could lead
to numeric dispersion and consequently to a low accuracy. 

It is possible
to make very fast simulations, in the present case for example, using $200$
radial zones the code take $\sim 10-30$ minutes to
run on a Pentium IV, 1.7 Mhz. 
Altough, to go further in the time evolution of the collapse could take much more time.
The total accuracy at the end of the simulation is $\sim 99.5\,\%$.
However, depending mostly on the choice of the viscocity parameter the accuracy could be worst. 
 
We must point out that the code only depart from excellent energy conservation only
when a trapped surface begins to form, so we think the precision obtained 
is quit satisfactory.

\vspace{.5cm}

\begin{figure}[h]
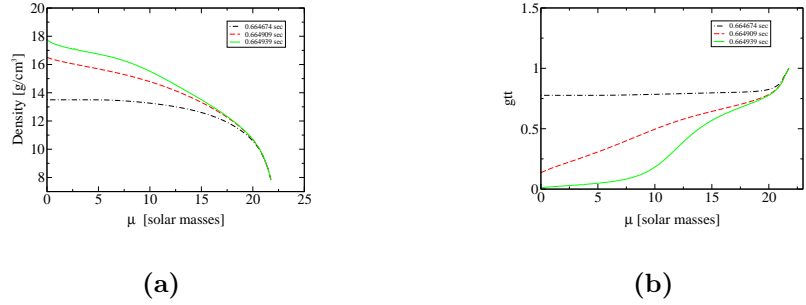

\begin{center}
$\begin{array}{c@{\hspace{1in}}c}
\multicolumn{1}{l}{\mbox{\bf (a)}} &
\multicolumn{1}{l}{\mbox{\bf (b)}} \\ [-0.53cm]
\epsfig{width=4cm,height=3cm, file=density2.eps} &
\epsfig{width=4cm,height=3cm, file=a.eps}
 \\ [0.4cm]
\mbox{\bf (a)} & \mbox{\bf (b)}
\end{array}$
\end{center}
\caption{The figures displays a time sequence of snapshots as a function of the Lagrangian coordinate 
$\mu$ for the density (Fig. a), and  for the metric coefficient 
$g_{tt}$ (Fig. b).}
\label{figtest-fig}
\end{figure}

\begin{figure}[h]
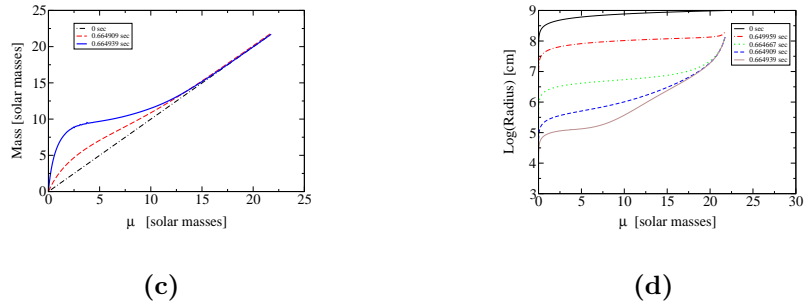

\begin{center}
$\begin{array}{c@{\hspace{1in}}c}
\multicolumn{1}{l}{\mbox{\bf (a)}} &
\multicolumn{1}{l}{\mbox{\bf (b)}} \\ [-0.53cm]
\epsfig{width=4cm,height=3cm, file=masas.eps} &
\epsfig{width=4cm,height=3cm, file=r.eps}
 \\ [0.4cm]
\mbox{\bf (c)} & \mbox{\bf (d)}
\end{array}$
\end{center}
\caption{Snapshots for the total mass or energy (Fig. c), and for the radius of the configuracion (Fig. d).}
\label{figtest-fig}
\end{figure}

\section*{Acknowledgments}
The authors acknowledge the brazilian agencies FAPESP,
CNpQ and CAPES for support. CRG specially acknowledge
FAPESP for additional support for the Tenth Marcel 
Grossmann meeting in Rio de Janeiro (2003).

\end{document}